# Thermodynamics of the Mg-B system: Implications for the deposition of $MgB_2$ thin films

Zi-Kui Liu and D. G. Schlom
Department of Materials Science and Engineering
The Pennsylvania State University
University Park, PA 16802
and
Qi Li and X. X. Xi
Department of Physics
The Pennsylvania State University
University Park, PA 16802

ABSTRACT

We have studied thermodynamics of the Mg-B system with the modeling technique CALPHAD using a computerized optimization procedure. Temperature-composition, pressure-composition, and pressure-temperature phase diagrams under different conditions are obtained. The results provide helpful insights into appropriate processing conditions for thin films of the superconducting phase, $MgB_2$, including the identification of the pressure/temperature region for adsorption-controlled growth. Due to the high volatility of Mg, $MgB_2$ is thermodynamically stable only under fairly high Mg overpressures for likely growth temperatures. This constraint places severe temperature constraints on deposition techniques employing high vacuum conditions.

The recent discovery of superconductivity in $MgB_2$ at 39 K has generated great interest [1,2]. $MgB_2$ has the highest $T_c$ known for non-oxide compounds and appears to become superconducting by the BCS mechanism [3]. Its coherent length is longer than those in high temperature cuprate superconductors, and its grain boundaries have a far less detrimental effect on superconducting current transport [4]. These properties hold tremendous promise for various large-scale and electronic applications. However, due to the high volatility of Mg, difficulties in fabricating $MgB_2$ thin films are anticipated. On the other hand, just as in the III-V and II-VI compound semiconductors, such volatility can greatly simplify composition control by enabling adsorption-controlled film growth [5]. It has been demonstrated for numerous materials containing a volatile constituent, an understanding of the thermodynamics of the system can help identify the appropriate growth region for these materials [6-10].

Although $MgB_2$ has been known and structurally characterized since the mid 1950's [11], no detailed thermodynamics information is available in the literature [12]. In this letter, we present a thermodynamic analysis of the Mg-B system with the thermodynamic modeling technique CALPHAD using a computerized optimization procedure. We have obtained temperature-composition, pressure-composition, and pressure-temperature phase diagrams under various conditions. We find that the $MgB_2$ phase is thermodynamically stable only under very high Mg vapor pressures. This condition is difficult to satisfy in deposition techniques employing high vacuum conditions. On the other hand, deposition techniques that can maintain a high flux of Mg may operate in a pressure-temperature window and exploit the automatic composition control benefit that accompanies adsorption-controlled growth.

In the CALPHAD approach, the Gibbs energies of individual phases in a system are constructed with models primarily based on the crystal structures of the phases [13]. For pure elements, the most commonly used model is suggested by the Scientific Group Thermodata Europe (SGTE) and the SGTE data have been compiled by Dinsdale [14]. By combining thermodynamic descriptions of unary systems with binary experimental data, thermodynamic descriptions of binary systems are then developed. In the Mg-B system, there are three intermediate compounds, $MgB_2$, $MgB_4$ and $MgB_7$, in addition to the gas, liquid, and solid (*hcp*) magnesium phases and the β-rhombohedral boron solid phase [15]. The Gibbs energy for the intermediate compounds is written using the two-sublattice model as

$$G_m^{MgB_b} = {}^\circ G_{Mg}^{solid} + b\,{}^\circ G_B^{solid} + \Delta G_f^{MgB_b} \qquad (1)$$

where $b$ is 2, 4 or 7 for the three intermediate phases, respectively, ${}^oG_{Mg}^{solid}$ and ${}^oG_B^{solid}$ are Gibbs energies for Mg and B solid, respectively, and $\Delta G_f^{MgB_b}$ is the Gibbs energy of formation for $MgB_b$. Using the experimentally-measured enthalpy of formation and estimated decomposition temperatures, $\Delta G_f^{MgB_b}$ is evaluated with the Thermo-Calc program [16] for $MgB_2$, $MgB_4$, and $MgB_7$, and the phase equilibria are then calculated.

In Fig. 1 the temperature-composition phase diagrams for the Mg-B system at (a) 1 atm, (b) 1 Torr, and (c) 1 mTorr are plotted. The labels "Solid", "Liquid", and "Gas" represent the Mg-rich solid, liquid, and gas phases, respectively. Our results at 1 atm pressure is consistent with the published Mg-B phase diagram [12]. Below 1545 °C and for the atomic Mg:B ratio, $x_{Mg}/x_B$, greater than 1:2, the $MgB_2$ phase coexists with the Mg-rich solid, liquid, and gas phases at various temepatures. Above 1545 °C $MgB_2$ decomposes into a mixture of $MgB_4$ and Mg vapor. If 1:4< $x_{Mg}/x_B$ <1:2 and the temperature is below 1545 °C, $MgB_2$ coexists with $MgB_4$. When the pressure is reduced to 1 Torr, the phase diagram changes dramatically. Since the pressure is lower than the triple-point pressure of Mg (650 °C, 2.93 Torr), the liquid phase of Mg completely disappears. The decomposition temperature of $MgB_2$ decreases to 912 °C. This temperature decreases further to 603 °C at the pressure of 1 mTorr. Evidently, the pressure has significant influence on the decomposition temperature of $MgB_2$, which can be very low thus severely limiting the thin film deposition temperature.

The kinetics of crystal growth requires that the film deposition takes place at sufficiently high temperatures. The optimum epitaxial growth temperature (in Kelvin) is typically about one half of the melting temperature, Tm, although the minimum temperature, which corresponds to the condition for polycrystalline film growth, can be much lower [17]. Thermodynamic calculation shows that $MgB_2$ melts congruently at 2430 °C (~ 2700 K) with pressure higher than 49000 Torr, therefore the optimum temperature for deposition of epitaxial $MgB_2$ films is around ~ 1080 °C (1350 K). In Fig. 2, the pressure-composition phase diagram at 850 °C is shown. At this temperature, $MgB_2$ is thermodynamically stable only above a Mg partial pressure of 340 mTorr. Below this pressure $MgB_2$ will decompose and only Mg vapor, $MgB_4$, $MgB_7$, or solid B can be obtained at 850 °C.

Figs. 1 and 2 illustrate the automatic composition control benefit that accompanies the adsorption-controlled growth. From the thermodynamics point of view, the film deposition condition should fall into a window where the stable phases are the desired phase $MgB_2$ and the gas phase (marked as "Gas + $MgB_2$"). From the figures we find that there are large composition windows in which Mg gas and $MgB_2$ coexist. As long as the Mg:B ratio is above the stoichiometric 1:2, any amount of extra Mg above the stoichiometry will be vaporized and the desired $MgB_2$ phase will result. The more critical requirement for controlling the stoichiometry is to avoid insufficient Mg supply, which will lead to $MgB_4$, $MgB_7$, or solid B phases.

The thermodynamic stability window for the $MgB_2$ film deposition is best illustrated by the pressure temperature phase diagram shown in Fig. 3. This phase diagram is essentially the same for all composition with $x_{Mg}/x_B \geq$ 1:2. For a given deposition temperature, one can find the necessary Mg vapor pressure range in the pressure-temperature window to keep the MgB2 phase thermodynamically stable. By converting the Mg vapor pressure to Mg flux from the deposition source using the formulism described in Ref. [18], the film deposition parameters can then be determined. For example, at the optimum temperature of about 1350 K the Mg vapor pressure has to be in the range of 10 Torr, which is highly impossible for many film deposition techniques. One can sacrifice epitaxy for phase stability by lowering the growth temperature, and the required Mg vapor pressure becomes more easily achievable. For example, one can lower the substrate temperature to below 550 °C (823 K) for molecular beam epitaxy, but this temperature is considerably lower than the optimum temperature for epitaxial growth.

In conclusion, the CALPHAD technique was used to develop a thermodynamic description of the Mg-B system. We find that $MgB_2$ is thermodynamically stable only under fairly high to very high Mg partial pressures at the temperature range appropriate for in-situ epitaxial growth, implying that a large Mg flux must be delivered from the deposition source. This requirement favors techniques like chemical vapor deposition and pulsed laser deposition, where large Mg flux can be maintained either uniformly or locally, over techniques like molecular beam epitaxy, where a large Mg flux is impractical. It should be pointed out that although the applicability of the equilibrium thermodynamics to thin film growth has been established in many material systems, the non-equilibrium nature of specific deposition techniques could change the details of the phase stability window.

This work is supported in part by ONR under grant No. N00014-00-1-0294 (XXX), by NSF under grant Nos. DMR-9983532 (ZKL),DMR-9876266 and DMR-9972973 (QL), and by DOE through grant DE-FG02-97ER45638 (DGS).

Figure Captions:

Figure 1. The temperature-composition phase diagrams of the Mg-B system under the pressures of (a) 1 atmosphere, (b) 1 Torr; and (c) 1 mTorr.

Figure 2. The pressure-composition phase diagram at 850$^{o}$C.

Figure 3. The pressure-temperature phase diagram for the Mg:B atomic ratio $x_{Mg}/x_B \geq 1/2$. The region of "Gas + $MgB_2$" represent the thermodynamic stability window for the deposition of $MgB_2$ thin films.

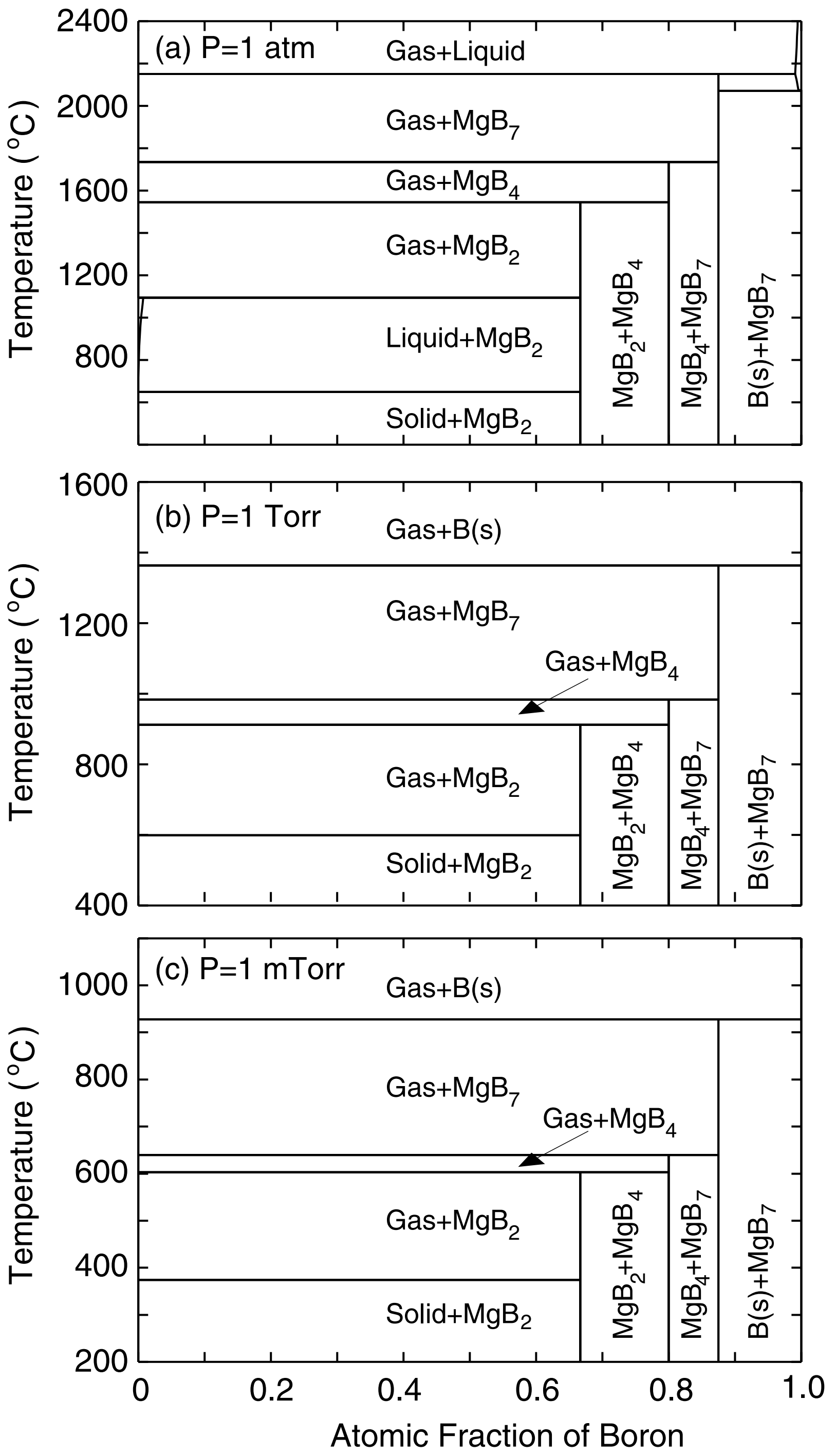

Fig.1 of 3
Liu et al.

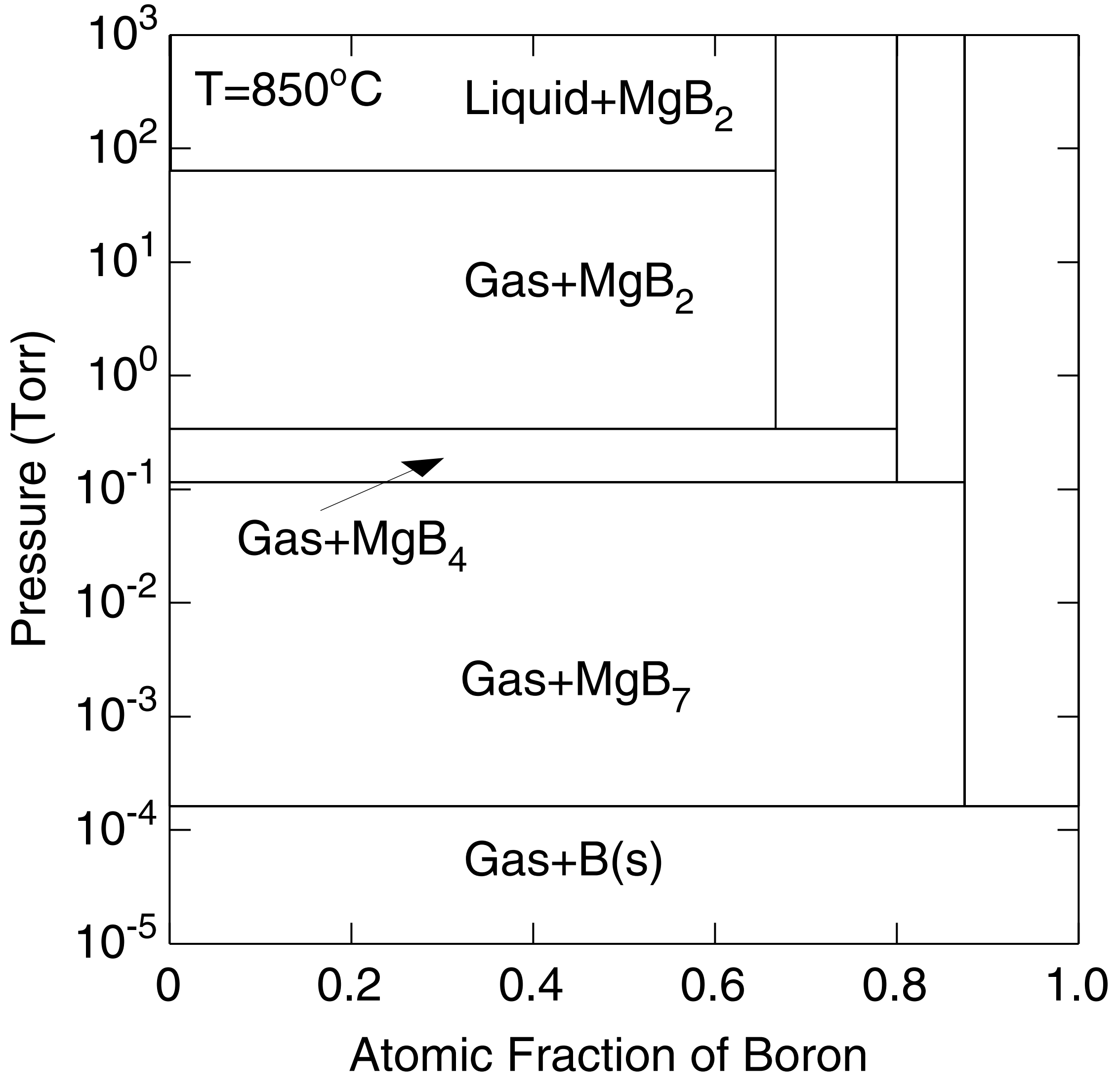

Fig.2 of 3
Liu et al.

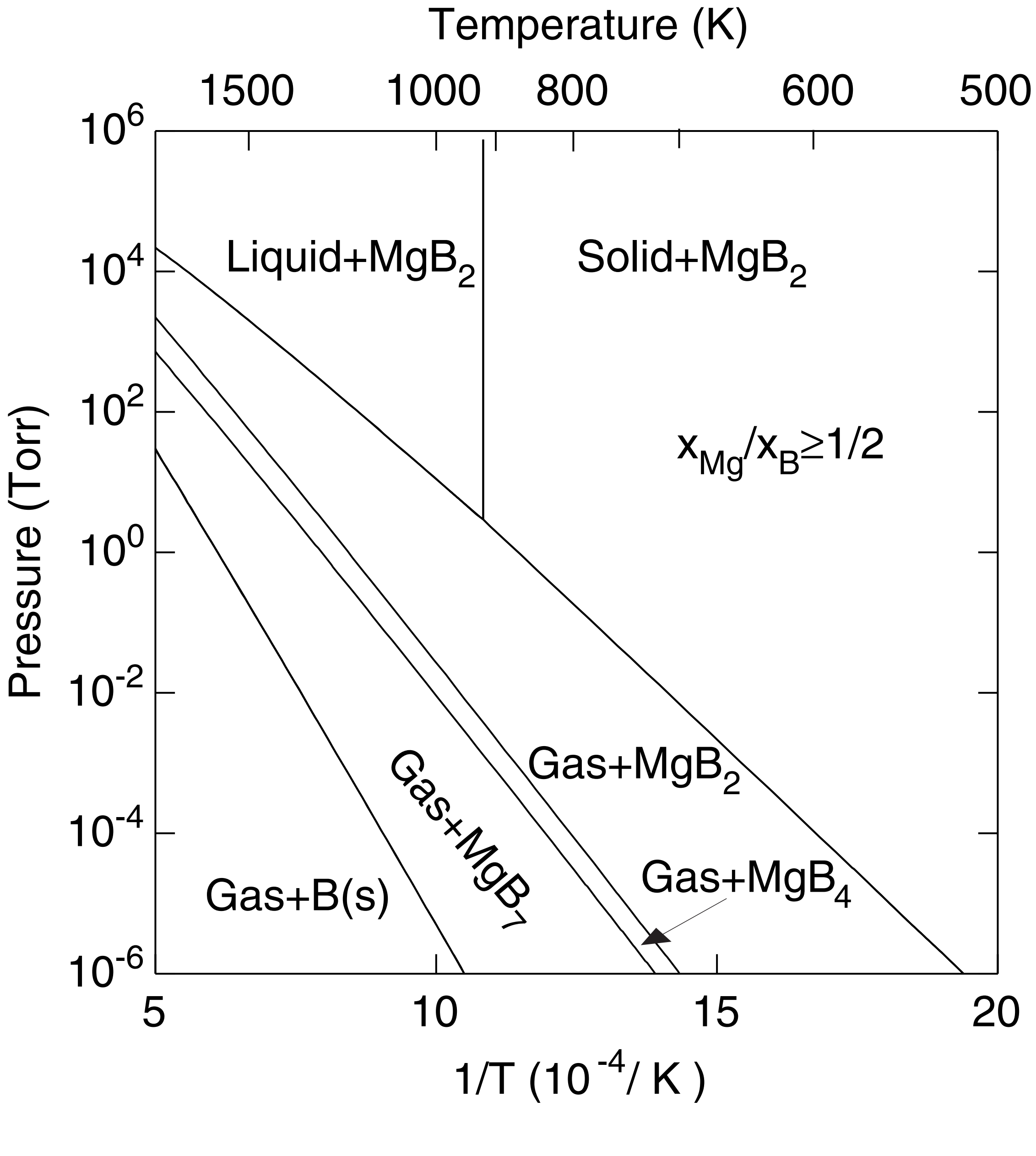

Fig.3 of 3
Liu et al.